\documentclass[usegraphicx,useAMS,usenatbib]{mn2e}

\newcommand{\kms}{\mbox{km\,s$^{-1}$}}
\def\kmsn{km\,${\rm s}^{-1}$}
\newcommand{\subs}[1]{$_{\rm #1}$}
\newcommand{\sups}[1]{$^{\rm #1}$}
\def\vsini{$\!${\em v\,}sin{\em i}}
\def\vsinis{$\!${\em v\,}sin{\em i} }
\def\ga{\mathrel{\hbox{\rlap{\hbox{\lower4pt\hbox{$\sim$}}}\hbox{$>$}}}}
\def\la{\mathrel{\hbox{\rlap{\hbox{\lower4pt\hbox{$\sim$}}}\hbox{$<$}}}}
\def\rdd{$\rm{rad\,{d}^{-1}}$}

\title[Magnetic fields and differential rotation on the pre-main sequence.]{Magnetic fields and differential rotation on the pre-main sequence III: The early-G star HD 106506}
\author[I. A. Waite et al.]{I.A.~Waite$^{1}$, S.C.~Marsden$^{2,3}$, B.D.~Carter$^{1}$,  R.~Hart$^{1}$, J.-F.~Donati$^{4}$, \and J.C.~Ram\'{i}rez V\'{e}lez $^{5,6}$, M.~Semel$^{4}$, and N.~Dunstone $^{7}$  \thanks{E-mail:waite@usq.edu.au} \\
${^1}$Faculty of Sciences, University of Southern Queensland, Toowoomba, 4350, Australia    \\
${^2}$Australian Astronomical Observatory, PO Box 296, Epping, Sydney, 1710, Australia             \\
${^3}$Centre for Astronomy, School of Engineering and Physical Sciences, James Cook University, Townsville, 4811, Australia \\
${^4}$LATT-UMR 5572, CNRS \& Univ. de Toulouse, 14 Av. E. Belin, F-31400 Toulouse, France \\
${^5}$LESIA, Observatoire de Paris-Meudon, F-92195 Meudon Cedex, France                    \\
${^6}$Instituto de Astronomia, Universidad Nacional Aut\'{o}noma de M\'{e}xico, 04510, Coyoacon, M\'{e}cixo D.F \\
${^7}$SUPA, School of Physics and Astronomy, University of St. Andrews, KY16 9SS, Scotland \\
}
\begin{document}

\date{}

\pagerange{\pageref{firstpage}--\pageref{lastpage}} \pubyear{2002}

\maketitle

\label{firstpage}

\begin{abstract}

We present photometry and spectropolarimetry of the pre-main sequence star HD~106506. A photometric rotational period of $\sim$1.416 $\pm$ 0.133 days has been derived using observations at Mount Kent Observatory (MKO). Spectropolarimetric data taken at the 3.9-m Anglo-Australian Telescope (AAT) were used to derive spot occupancy and magnetic maps of the star through the technique of Zeeman Doppler imaging (ZDI). The resulting brightness maps indicate that HD~106506 displays photospheric spots at all latitudes including a predominant polar spot. Azimuthal and radial magnetic images of this star have been derived, and a significant azimuthal magnetic field is indicated, in line with other active young stars. A solar-like differential rotation law was incorporated into the imaging process. Using Stokes $\it{I}$ information the equatorial rotation rate, $\Omega$\subs{eq}, was found to be 4.54 $\pm $0.01 \rdd, with a photospheric shear $\delta\Omega$ of $0.21_{-0.03}^{+0.02}$  \rdd. This equates to an equatorial rotation period of $\sim$1.39 $\pm$ 0.01 days, with the equatorial region lapping the poles every $\sim$ $30_{-3}^{+5}$ days. Using the magnetic features, the equatorial rotation rate, $\Omega$\subs{eq}, was found to be 4.51 $\pm$ 0.01 \rdd, with a photospheric shear $\delta\Omega$ of 0.24 $\pm$ 0.03 \rdd. This differential rotation is approximately 4 times that observed on the Sun.

\end{abstract}


\section{Introduction}

Young solar-type F, G, and K stars provide proxies for the Sun's early evolution and insight into its intense pre-main sequence and zero-age main sequence magnetic activity. In this regard, observations of starspots and surface magnetic fields are important, as they provide the empirical basis for understanding the star's magnetic dynamo and its internal structure. Differential rotation is one of the primary drivers of the dynamo process that produces magnetic fields within the Sun. In the solar convection zone, differential rotation is constant down to the base of this zone in the region known as the tachocline. Strong shear forces are formed at the interface between the solid-body rotation of the radiative zone and the differentially rotating convective layer. This interaction converts the poloidal magnetic field to a toroidal field. This effect is known as the ``$\Omega$-effect''. However, on young solar-type stars, the operation of the magnetic dynamo process may be occurring throughout the convective zone itself. This is known as a distributed dynamo \citep{Brandenburg89,Moss95}.

\citet{Barnes05} have investigated the link between differential rotation and effective temperature, and hence convective turnover time. They were able to fit an empirical power law to recent observations of a small sample of stars. As the temperature increases, so does the rotational shear. A small number of late F-/early G-type stars have been investigated using Doppler imaging (DI) to determine their differential rotation. HD 171488 (V889 Her) \citep{Marsden06}, HD 307938 (R58 in IC 2602) \citep{Marsden05}, HR 1817 \citep{Mengel05} and LQ Lup (RX J1508.6-4423) \citep{Donati00} have had their differential rotation determined.

Zeeman Doppler imaging (ZDI) \citep{Donati03a} enables the study of stellar magnetic fields using spectropolarimetry and offers a window into the underlying magnetic dynamo. ZDI requires high signal-to-noise data as the polarisation signature is typically less than 0.1\% of the total light intensity \citep{Donati97}. ZDI only measures the large-scale magnetic fields as small-scale magnetic fields cannot be recovered as the positive and negative magnetic fields within the resolution element are likely to be cancelled out. ZDI has been achieved on a small sample of single, G-type stars such as HD 171488 (V889 Her) \citep{Marsden06} and HD 141943 \citep{Marsden10a, Marsden10b}.

HD 106506 is a suitable target for DI and ZDI. \citet{Henry96} identified this star as active while \citet{Mason98} and \citet{Cutispoto02} deduced this star as being single. The high resolution spectroscopic survey of \citet{Waite05} confirmed this star as a young, rapidly rotating solar-type star. They measured a rotation rate of \vsinis $\sim$ 80 \kmsn, strong H$\alpha$ chromospheric activity (log$R$\sups{\prime}\subs{H\alpha} $\sim$4.2), and deformation of the spectral line profiles indicating the presence of large starspots. 

The aim of this investigation is to map the surface of this star with a view to measuring the rotational shear on its surface and to measure and to map the magnetic field structure on the surface of this star. This is the third paper in a series that investigates the magnetic field topologies of young sun-like stars, with HD~141943 being the subject of two earlier papers \citep{Marsden10a,Marsden10b}.
  
\begin{figure}
\begin{center}
\includegraphics[scale=0.522, angle=0]{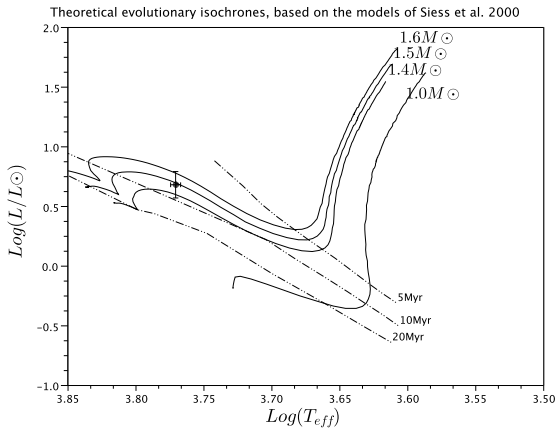}
\caption{The evolutionary status of HD 106506. This is based on the theoretical models of \citet{Siess00}}
\label{evolution} 
\end{center}
\end{figure}

\section[Observations and Analysis]{Observations and Analysis}

Spectropolarimetry of HD 106506 was obtained using the 3.9-m Anglo-Australian Telescope (AAT) over an 11 night period from the 30th of March - 9th of April, 2007. Near-simultaneous BVR photometric data were obtained using the University of Louisville's 50cm telescope at the University of Southern Queensland's Mount Kent Observatory (MKO). MKO is located in Southern Queensland, approximately 160km south-west of Brisbane. 

\subsection{BVR Photometry at the MKO}

The University of Louisville 50cm telescope at MKO is a Corrected Dall-Kirkham design manufactured by Planewave Instruments. BVR photometric observations of HD~106506 were taken through the standard Cousins B, V and R filters using an SBIG STL-630E with a Kodak KAP-6303E CCD chip that has an array size of 3060 x 2040, 9 $\mu$m square pixels giving a plate scale of approximately 0.54 arc seconds per pixel \citep{mko}. Table \ref{photometry_log} shows the journal of observations while a more detailed analysis of this photometric study of HD~106506 has been included in appendix A. 

\begin{table}
\begin{center}
\caption{The Journal of photometric observations of HD~106506 showing the date and filters used. The E5-Region (Graham, 1982) was observed on the 11th April, 2007.} 
\label{photometry_log}
\begin{tabular}{lccc}
\hline
UT Date    & Filters Used  & UT begin   & UT end  \\
\hline
2007 April 01    & B,V,R   & 09:12:11   & 19:03:53 \\
2007 April 02    & B,V,R   & 11:09:03   & 17:57:21 \\
2007 April 03    & B,V,R   & 09:48:02   & 18:10:43 \\
2007 April 04    & B,V,R   & 14:33:42   & 17:59:18 \\
2007 April 05    & B,V,R   & 11:06:02   & 17:14:08 \\
2007 April 06    & B,V,R   & 09:16:22   & 18:19:50 \\
2007 April 07    & B,V,R   & 09:13:45   & 18:13:12 \\
2007 April 08    & B,V,R   & 09:06:57   & 18:17:01 \\
2007 April 11    & B,V,R   & 10:10:43   & 13:26:59 \\
\hline
\end{tabular}
\end{center}
\end{table}

\subsection{High Resolution Spectropolarimetric Observations from the AAT}

High resolution spectropolarimetric data were obtained from the AAT using the University College of London \'{E}chelle Spectrograph (UCLES) and the Semel Polarimeter (SEMPOL) \citep{Semel89,Semel93,Donati03a}. The detector used was the deep depletion EEV2 CCD with 2048 x 4096 13.5$\mu$m square pixels. UCLES was used with a 31.6 gr/mm grating covering 46 orders (orders \# 84 to \# 129). The central wavelength was 522.002 nm with full wavelength coverage from 437.71 nm to 681.56 nm. The dispersion of $\sim$0.004958 nm at order \# 129 gave a resolution of approximately 71000. Observations in the circular polarization signatures (Stokes {\it V}) consists of a sequence of four exposures. After each of the exposures, the half-wave Fresnel Rhomb of the SEMPOL polarimeter is rotated between +45$^o$ and -45$^o$ so as to remove instrumental polarization signals from the telescope and the polarimeter \citep{Donati03a}. A journal of observations is shown in Table \ref{spectroscopy_log}.

\begin{table}
\begin{center}
\caption{The Journal of spectropolarimetric observations of HD~106506 using the 3.9-m AAT.} 
\label{spectroscopy_log}
\begin{tabular}{lccc}
\hline
UT Date &  UT begin          & UT end ${^1}$     & Number          \\ 
        &                    &            	 & of Sequences ${^2}$ \\
\hline
2007 March 30   & 09:02:56 & 13:41:16 & 5   \\
2007 March 31   & 08:51:37 & 13:36:05 & 5   \\
2007 April 01   & 08:51:37 & 13:36:05 & 5   \\
2007 April 02   & 08:55:31 & 13:38:07 & 5   \\
2007 April 03   & 09:26:01 & 16:25:01 & 3   \\
2007 April 04   & 08:57:07 & 12:42:40 & 2   \\
2007 April 05   & 10:13:25 & 14:06:13 & 4   \\
2007 April 06   & 09:08:53 & 12:07:27 & 2   \\
2007 April 07   & 08:49:55 & 13:28:11 & 5   \\
2007 April 08   & 10:31:20 & 11:15:43 & 1   \\
2007 April 09   & 09:35:14 & 11:53:58 & 2   \\

\hline
\end{tabular}
\end{center}
{${^1}$ The UT end is the time when the final exposure for HD 106506 was completed.}\\
{${^2}$ Each sequence of 4x600 seconds exposures. After each exposure, the half-wave Fresnel Rhomb of the SEMPOL polarimeter is rotated between +45$^o$ and -45$^o$ so as to remove instrumental polarization signals from the telescope and the polarimeter.}
\end{table}

\subsubsection{Spectropolarimetric Analysis}
The initial data reduction was completed using the {\small ESpRIT} (\'{E}chelle Spectra Reduction: an Interactive Tool) software package \citep{Donati97,Donati03a}.  Preliminary processing involved removing the bias and using a nightly master flat combining typically 20 flat field exposures. Each stellar spectrum was extracted and wavelength calibrated against a Thorium-Argon lamp. The mean pixel resolution for the AAT spectra was determined to be 1.689 \kms pixel$^{-1}$. 

After using {\small ESpRIT}, Least Squares Deconvolution (LSD) was applied to the reduced spectra. LSD combines the information from many spectral lines to produce a single line profile, thereby providing an enormous multiplex gain in the signal-to-noise (S/N) \citep{Donati97}. The average S/N for a typical Stokes $\it{I}$ LSD profile was $\sim$1100 whereas the maximum S/N for the corresponding central order was only $\sim$50. A G2 line mask created from the Kurucz atomic database and ATLAS9 atmospheric models \citep{Kurucz93} was used during the LSD process. 

In order to correct for the minor instrumental shifts in wavelength space, each spectrum was shifted to match the Stokes {\it I} LSD profile of the telluric lines contained in the spectra, as was done by \citet{Donati03a} and \citet{Marsden06}. Further information on LSD can be found in \citet{Donati97}.

\section[Photospheric and Chromospheric Features]{Photospheric and Chromospheric Features}

The brightness and magnetic images of HD 106506 were created using the ZDI code of \citet{Brown91} and \citet{DonatiBrown97}. It utilizes the time-series Stokes $\it{I}$ and V LSD profiles as inputs for the inversion process to produce the brightness and magnetic images. This is an ill-posed problem as there are an infinite number of possible solutions that can be found. The ZDI code uses the maximum-entropy minimisation techniques of \citet{Skilling84} to produce an image with the mimimum amount of information required to produce the observed variations in the LSD profile. 

\subsection[Fundamental Parameters of HD 106506]{Fundamental Parameters of HD 106506}

HD 106506 is a G1V star \citep{Torres06} with a median visual magnitude of 8.5617 \citep{Van_Leeuwen07}, a B-V value of 0.605 and a V-I value of 0.67 (based on the transformed Hipparcos and Tycho data, \citet{ESA97}). It has a trigonometric parallax of 7.96 $\pm$ 0.88 \citep{Perryman97} giving a distance of 126 $\pm$ 14pc or 410 $\pm$ 46lyr. Using this V-I value, the temperature of HD 106506 is approximately 5870 $\pm$ 45K. This was determined using the formulation of Bessell, Castelli \& Plez (1998). The absolute magnitude is 3.053 $\pm$ 0.244. Using the bolometric corrections listed in \citet{Bessell98} and the formulation contained within that paper, HD 106506 is estimated to be 2.15 $\pm$ 0.26 $R_{\odot}$. The resulting luminosity of HD 106506 is 4.81 $\pm$ 1.2 $L_{\odot}$. Figure \ref{evolution} shows the theoretical tracks of \citet{Siess00}, from which it may be deduced that the star is less than 10 Myr of age and has a mass of between 1.4 and 1.6 $M_{\odot}$. Table \ref{parameters} shows the initial parameters that were used in the analysis and the production of the Stokes $\it{I}$ (intensity) and Stokes $\it{V}$ magnetic maps. 

\begin{table}
\begin{center}
\caption{The parameters used to produce the maximum-entropy image reconstruction of HD 106506, including those associated with surface differential rotation.} 
\label{parameters}
\begin{tabular}{ll}
\hline
Parameter         			& Value                   \\
\hline
Photometric Period 			& 1.416 $\pm$ 0.133 days  \\
Equatorial Period (Using Stokes $\it{I}$) 	& 1.39 $\pm$ 0.01 days   \\
Inclination Angle 			& 65 $\pm$ 5$^{o}$        \\
Projected Rotational Velocity, \vsini 	& 79.5 $\pm$ 0.5 \kmsn    \\
Unspotted Apparent Visual magnitude     & 8.38                    \\
Photospheric Temperature, T\subs{phot} 	& 5900 $\pm$ 50 K         \\
Spot Temperature, T\subs{spot} 		& 4000 $\pm$ 50 K         \\
Radial Velocity, v\subs{rad} 		& 13.1 $\pm$ 0.1 \kmsn    \\

Stokes $\it{I}$: $\Omega$\subs{eq} 		& 4.54 $\pm$ 0.01 \rdd  \\
Stokes $\it{I}$: d$\Omega$			& $0.21_{-0.03}^{+0.02}$  \rdd  \\
Stokes $\it{V}$: $\Omega$\subs{eq}		& 4.51 $\pm$ 0.01  \rdd  \\
Stokes $\it{V}$: d$\Omega$			& 0.24 $\pm$ 0.03  \rdd  \\

\hline
\end{tabular}
\end{center}
\end{table}

\subsection{Image Reconstruction: Stokes $\it{I}$ Doppler imaging map and surface differential rotation }

In order to produce the brightness image of HD 106506, a two-temperature model was used \citep{Cameron92}, assuming one temperature was for the quiet photosphere,T\subs{phot}, while the other temperature is for the cooler spots, T\subs{spot}. Synthetic gaussian profiles were used to represent both the spot temperature and the quiet photosphere temperature. \citet{Unruh95} showed that there is little difference in the resulting maps when using synthetic gaussian profiles when compared with using profiles that were taken of slowly rotating standard stars. A number of authors have followed this formulation such as \citet{Petit02,Petit04} and \citet{Marsden05,Marsden06}. 

\begin{figure*}
\begin{center}
\includegraphics[scale=0.70, angle=-90.0]{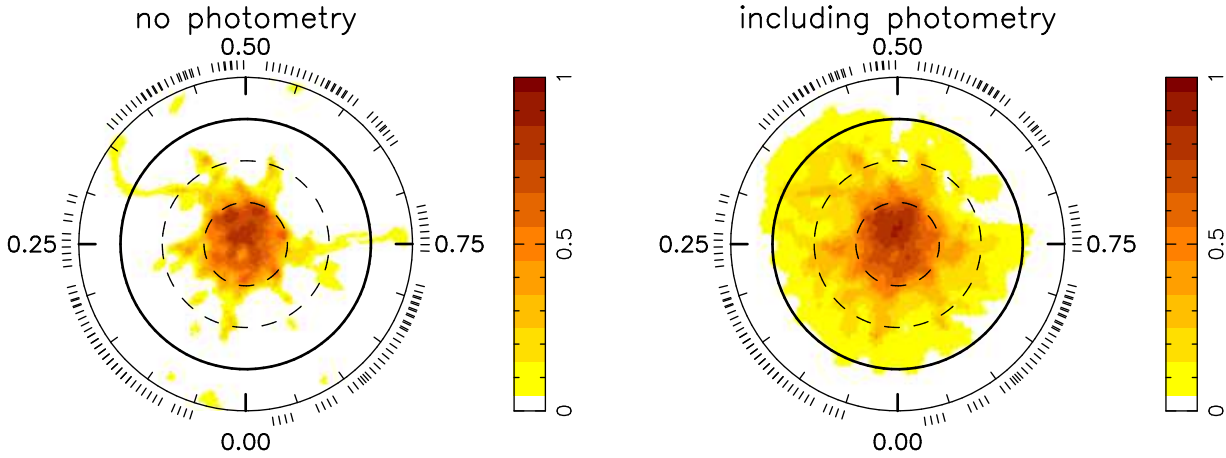}
\caption{(a) The figure on the left is the maximum entropy map that was generated entirely from the AAT data, with a spot filling factor of 0.062. (b) The figure on the right is a similar map but including near simultaneous photometry from the MKO. The spot filling factor was 0.127. The $\chi{^2}$ value was set to 0.5 in the modelling process. The images are polar projections extending down to a latitude of -30$^{o}$. The bold line denotes the equator and the dashed lines are +30$^{o}$ and +60$^{o}$ latitude parallels. The radial ticks indicate the phases at which this star was observed spectroscopically. }
\label{Photometry_incl}
\end{center}
\end{figure*}

The imaging code was used to establish the values of a number of basic parameters, including the star's inclination angle, projected rotational velocity, \vsini, and radial velocity, v\subs{rad}. This was achieved by systematically varying each parameter in order to minimise the $\chi{^2}$ value (eg see \citet{Jeffers08,Marsden05}). These key stellar parameters were determined sequentially, first v\subs{rad}, then \vsini, and finally the inclination angle of the star. This sequence was repeated each time additional parameters were modified as a result of the imaging process. The full set of parameters that gave the minimum $\chi{^2}$ value are shown in Table \ref{parameters} and were adopted when producing the final Doppler imaging map in Figure \ref{Photometry_incl}(a). By incorporating the photometry from MKO into the imaging process, T\subs{phot} and T\subs{spot} were determined by minimising the deviations between the measured and modelled data in both the V- and R-bands. The effective photospheric temperature was determined to be 5900 $\pm$ 50K, which is consistent with that calculated based on the photometric colours of the star and using the bolometric corrections of \citet{Bessell98} within the error bars of both determinations. The spot temperature was determined to be 4000 $\pm$ 50K. The photospheric-spot temperature difference of 1900K for this early G-dwarf is consistent with measurements made by other authors such as \citet{Marsden10a,Marsden10b,Marsden05} and supports the relationship developed by \citet{Berdyugina05}. The intensity map with the photometry included is shown in Figure \ref{Photometry_incl}(b).

There have been a number of different ways in which differential rotation have been measured. \citet{Reiners03} have used a Fourier transform method to derive the parameters for rapidly rotating F-type / early G-type inactive stars. They produce a deconvolved line profile from the stellar spectra and then determine the ratio of the second and first zero of the resulting Fourier transform. This ratio is a measure of differential rotation on the star. However, for active young stars with significant asymmetry within the line profiles, this technique is not as effective. The differential rotation of a star can also be estimated by tracking spot features at different latitudes on the surface of the star. A solar-like differential rotation law, as defined in equation~\ref{DR}, has been applied in a number of solar-type stars (i.e.\citet{Donati03a, Marsden05, Marsden06}). 
\begin{equation}
	\label{DR}	
	\Omega(\theta) = \Omega_{eq} - d\Omega sin^{2} \theta 
\end{equation}
where $\Omega(\theta)$ is the rotation rate at latitude $\theta$ in \rdd, $\Omega$\subs{eq} is equatorial rotation rate, and $d\Omega$ is the rotational shear between the equator and the pole. In this paper the differential rotation of HD~106506 was measured using a technique that systematically adjusts the differential rotation parameters, $\Omega$\subs{eq} and d$\Omega$ and determines the best fit to the data, by minimising the difference between the data and the model fits using $\chi^{2}$ minimisation techniques. This technique has been successfully applied to a number of late F-/early-G stars (i.e. \citet{Petit02, Petit04, Barnes05, Marsden06}).  


Initially, a fixed spot filling factor of 0.062, as determined from the reconstructed map, was used and $\Omega$\subs{eq} and $d\Omega$ were considered to be free parameters. The magnitude of the differential rotation was then determined by fitting a paraboloid to the reduced $\chi^{2}$ values. The reduced $\chi^{2}$ values for the various $\Omega$\subs{eq} and $d\Omega$ is shown in Figure \ref{DR_1}. The associated error ellipse, as shown in Figure \ref{DR_1} was determined by varying both the spot filling factor and the inclination angle. Both were changed by $\pm$ 10$\%$. When varying the inclination angle, the associated multiplications factors were re-determined and the resulting spot filling factor as found to match the new angle by minimising the $\chi^{2}$ value.

\begin{figure*}
\begin{center}
\includegraphics[scale=0.70, angle=-90.0]{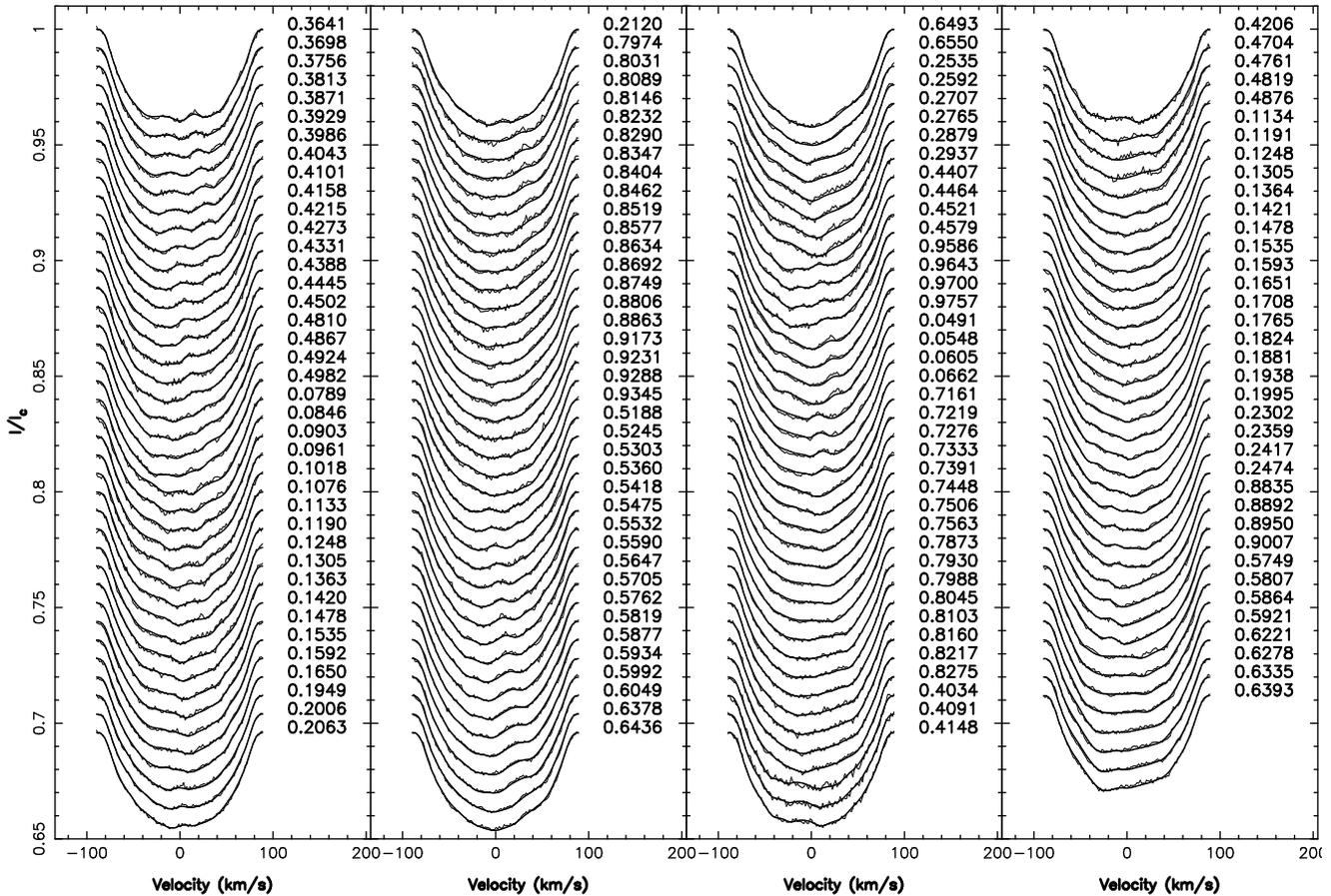}
\caption{The maximum-entropy fits to the LSD profiles for HD 106506 with both photometry and surface differential rotation incorporated into the analysis. The thick lines represent the modelled lines produced by the Doppler imaging process whereas the thin lines represent the actual observed LSD profiles. Each successive profile has been shifted down by 0.005 for graphical purposes. The rotational phases at which the observations took place are indicated to the right of each profile.}
\label{fit_2}
\end{center}
\end{figure*}

\begin{figure}
\begin{center}
\includegraphics[scale=0.50, angle=0.0]{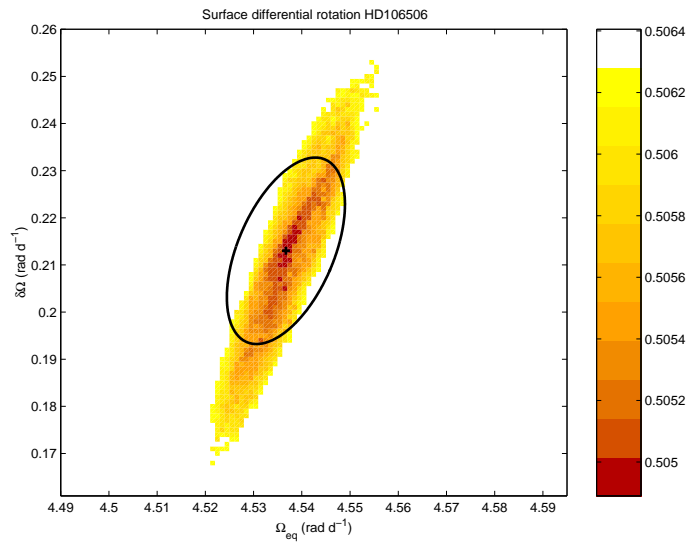}
\caption{Surface differential rotation $\chi^{2}$ minimisation for HD 106506. The image shows the reduced $\chi^{2}$ values from the maximum-entropy Doppler imaging code for a fixed spot coverage of 0.062 and an inclination angle of 65$^{o}$. The darker regions correspond to lower $\chi^{2}$ values. The image scale projects $\pm$ 7$\sigma$ on to the axes in both $\Omega$\subs{eq} and $d\Omega$. The ellipse superimposed demonstrates the error for the value of the differential rotation and was determined by varying both the star's inclination and spot filling factors by $\pm$ 10\% and determining $\Omega$\subs{eq} and $d\Omega$.}
\label{DR_1}
\end{center}
\end{figure}

The minimum value for $\Omega$\subs{eq} was determined to be 4.54 $\pm$ 0.01 \rdd and $d\Omega$ = $0.21_{-0.03}^{+0.02}$ \rdd. The errors for the value of the differential rotation were determined by varying both the star's inclination and spot filling factors by $\pm$ 10\% and determining $\Omega$\subs{eq} and $d\Omega$ respectively. By incorporating the differential rotation into the maximum-entropy image reconstruction, the resulting map is shown in Figure \ref{Photometry_incl}, along with the model fits to the LSD profiles in Figure \ref{fit_2}.

Figure \ref{Photometry_incl} shows, on the left, the resulting maximum-entropy image reconstructed map generated using the spectroscopic data taken at the AAT. On the right of this figure is the reconstructed image created using both the spectroscopic data and the near-simultaneous photometric data taken at the MKO. The photometry data has been able to enhance the lower-latitude features. In each map, an equatorial period of 1.39  days has been used, which was established during the differential rotation analysis and the epoch was set to 2454194.93642361, which was the middle of the observation run.

\begin{figure}
\begin{center}
\includegraphics[scale=0.40, angle=0.0]{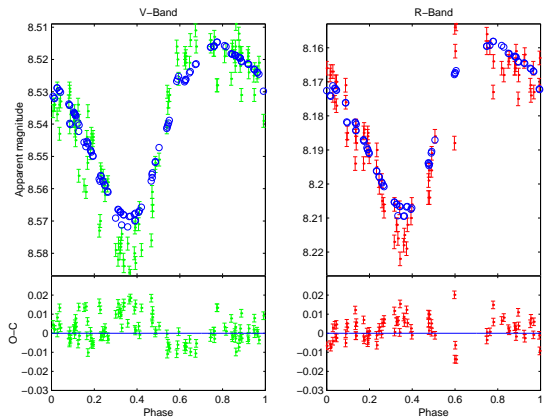}
\caption{The photometric light curves for both the V filter and R filter (the points with error bars) coupled with the maximum-entropy fits (circles) for HD 106506. The bottom graph in each panel shows the Observed-Calculated values.}
\label{OC_Graph}
\end{center}
\end{figure}

\begin{figure}
\begin{center}
\includegraphics[scale=0.38, angle=0.0]{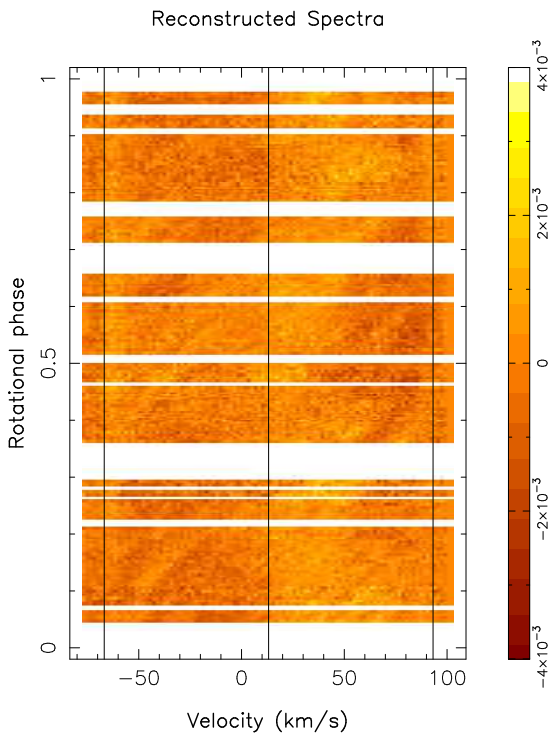}
\caption{Dynamic spectra of the residuals between the observed-modelled LSD profiles, including differential rotation. The spectra is centred on the radial velocity of HD 106506. The three lines on the image shows the centre of the line profile (centre line) and the rotational broadening of the line profile (outer lines), which is 79.5 \kms.}
\label{Residuals}
\end{center}
\end{figure}

Figure \ref{OC_Graph} shows the V and R light curves coupled with the maximum-entropy fits that were used to produce the enhanced DI map. Figure \ref{fit_2} shows the model profiles overlayed on the measured LSD profiles. The model profiles include the differential rotation parameters. The dynamic spectrum of the residuals between the observed and modelled profiles is shown in Figure \ref{Residuals}. It appears that the Doppler imaging process, along with incorporating the differential rotation parameters, has removed the majority of the large scale features with only a small amount of structure still apparent in this dynamic spectrum.

\subsection{Zeeman Doppler imaging: magnetic features on the surface of HD~106506}

The modelling strategy of \citet{DonatiBrown97} was used to construct the magnetic field topology on HD~106506. This involved utilising the spherical harmonic expansions of the surface magnetic field as implemented by \citet{Donati06}. The spherical harmonic expansion {$\it l_{max}$} = 40 was selected as this was the minimum value where any further increase resulted in no further changes in the magnetic topologies. 

Initially, a poloidal plus toroidal field was assumed, then the images were created by fitting the data to within the noise level and resulted in a mean field strength of 68.84 G. A similar approach to measuring the differential rotation parameters as was used for the Stokes $\it{I}$ (brightness) features. When considering these magnetic features, the equatorial rotation rate $\Omega$\subs{eq} was 4.51 $\pm$ 0.01 \rdd, with a photospheric shear $\delta\Omega$ of 0.24 $\pm$ 0.03 \rdd. This implies that the equatorial rotation period was 1.38 days with a shear approximately 4 times that of the solar value.

\begin{figure*}
\begin{center}
\includegraphics[scale=0.75, angle=-90.0]{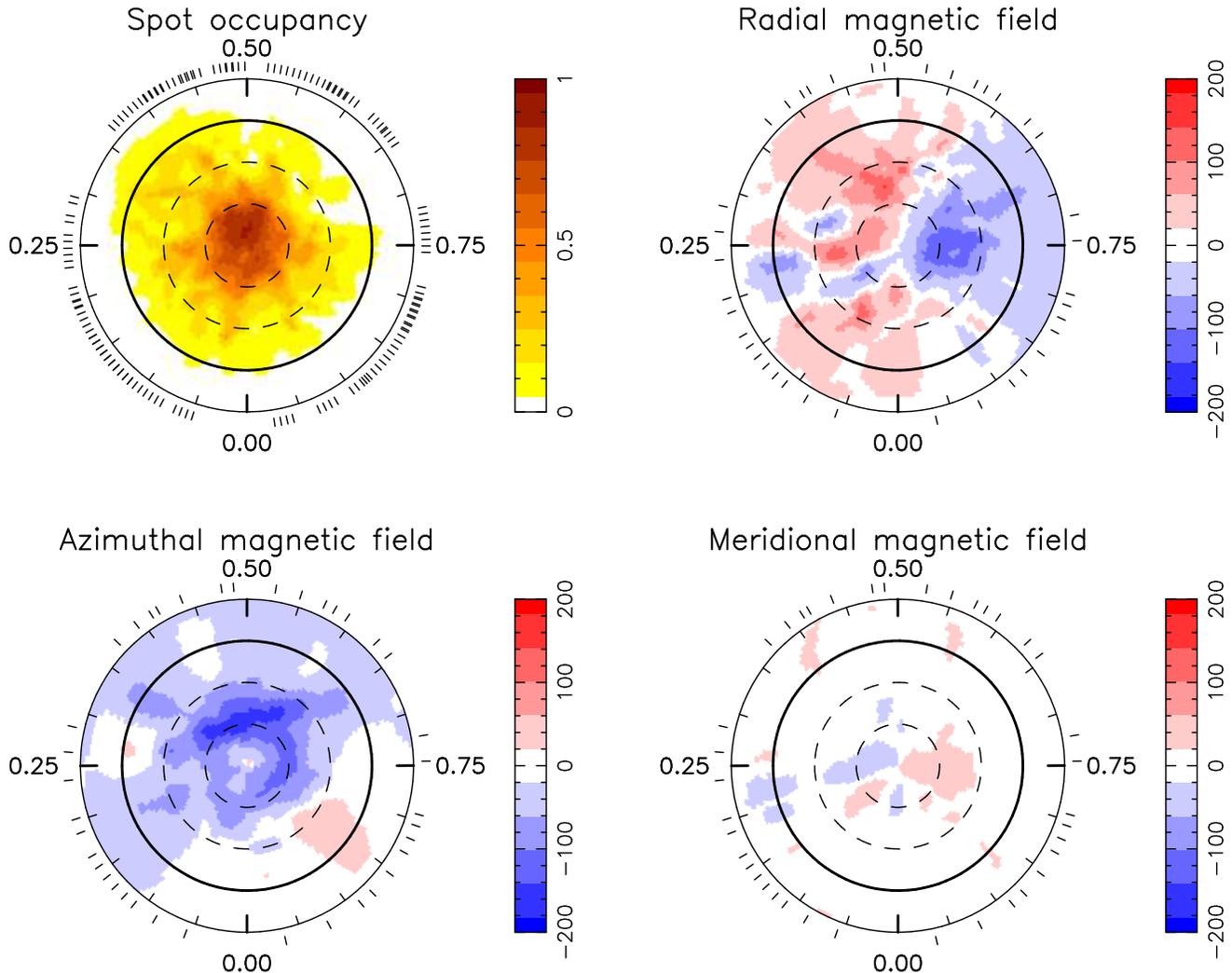}
\caption{The maximum-entropy brightness and magnetic image reconstructions for HD~106506. These maps are polar projections extending down to -30${^o}$. The bold lines denote the equator and the dashed lines are +30${^o}$ and +60${^o}$ latitude parallels. The radial ticks indicate the phases at which this star was observed spectroscopically. The scale of the magnetic images are in Gauss. The brightness image (top left-hand image) has a spot filling factor of $\sim$0.127 (or 12.7\%) and is a combination of both the spectroscopic and photometric data. The magnetic images have a field modulus of 68.84G. Differential rotation, as measured using the Stokes $\it{V}$ profiles, has been incorporated into the analysis.}
\label{mapmag}
\end{center}
\end{figure*}

\begin{figure*}
\begin{center}
\includegraphics[scale=0.95, angle=00.0]{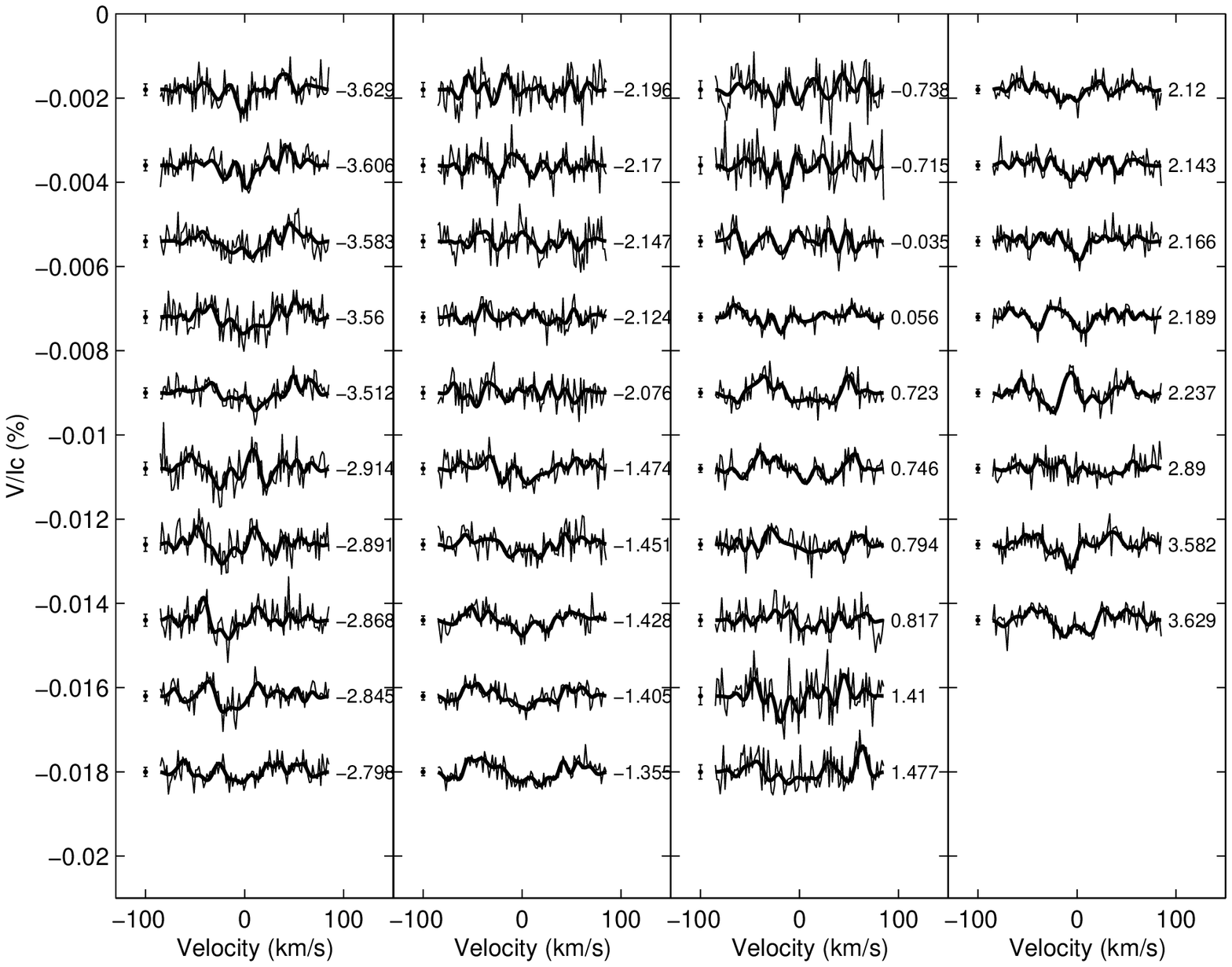}
\caption{The maximum-entropy fits to the Stokes $\it{V}$ LSD profiles for HD 106506 with surface differential rotation, as measured using the Stokes $\it{V}$ profiles, incorporated into the analysis. The thick lines represent the modelled lines produced by the Zeeman Doppler imaging process whereas the thin lines represent the actual observed LSD profiles. Each successive profile has been shifted down by 0.002 for graphical purposes. The error bar on the left of each profile is plotted to $\pm$ 0.5 $\sigma$. The rotational phases, in terms of the number of rotational cycles, at which the observations took place are indicated to the right of each profile.}
\label{mag_fits}
\end{center}
\end{figure*}

The reconstructed magnetic fields are shown in Figure \ref{mapmag}. The fits to the Stokes $\it{V}$ LSD profiles are given in Figure \ref{mag_fits}.  

\subsection{Chromospheric Features: Prominences}

The H$\alpha$ line has been extensively used as a indicator of chromospheric activity (eg \citet{Thatcher93}).  

\begin{figure}
\begin{center}
\includegraphics[scale=0.38, angle=0.0]{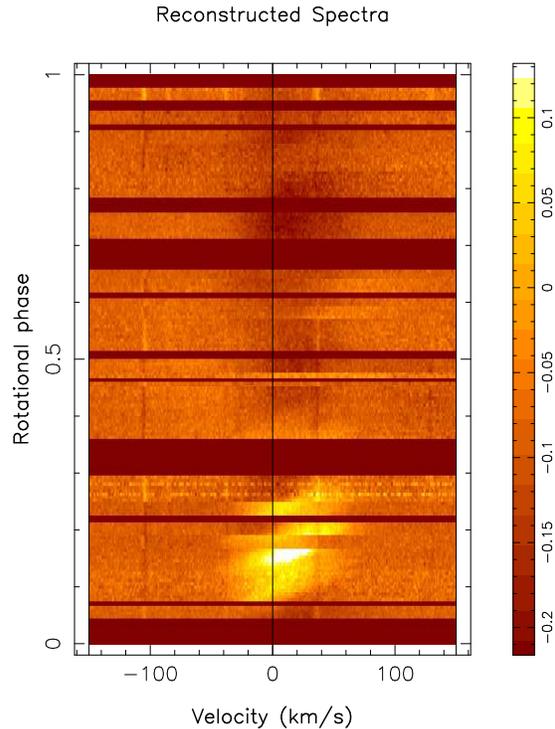}
\caption{This figure shows a chromospheric features observed on HD~106506. The chromosphere is more dense at phase of $\sim$ 0.15. There also appears to be an area at $\sim$ 0.75 where the chromosphere is less dense.}.
\label{chromo1}
\end{center}
\end{figure}

A mean H$\alpha$ spectral line for HD 106506 was generated from all the H$\alpha$ profiles obtained during the observing run. Each individual spectrum was then subtracted from the mean H$\alpha$ line, and dynamic spectra were then produced. Chromospheric features were observed at phases $\sim$0.15 and $\sim$0.75. This is shown in Figure \ref{chromo1}. 

\section{Discussion}
\subsection{Spot Features and Surface Differential Rotation}

\begin{figure}
\begin{center}
\includegraphics[scale=0.45, angle=0.0]{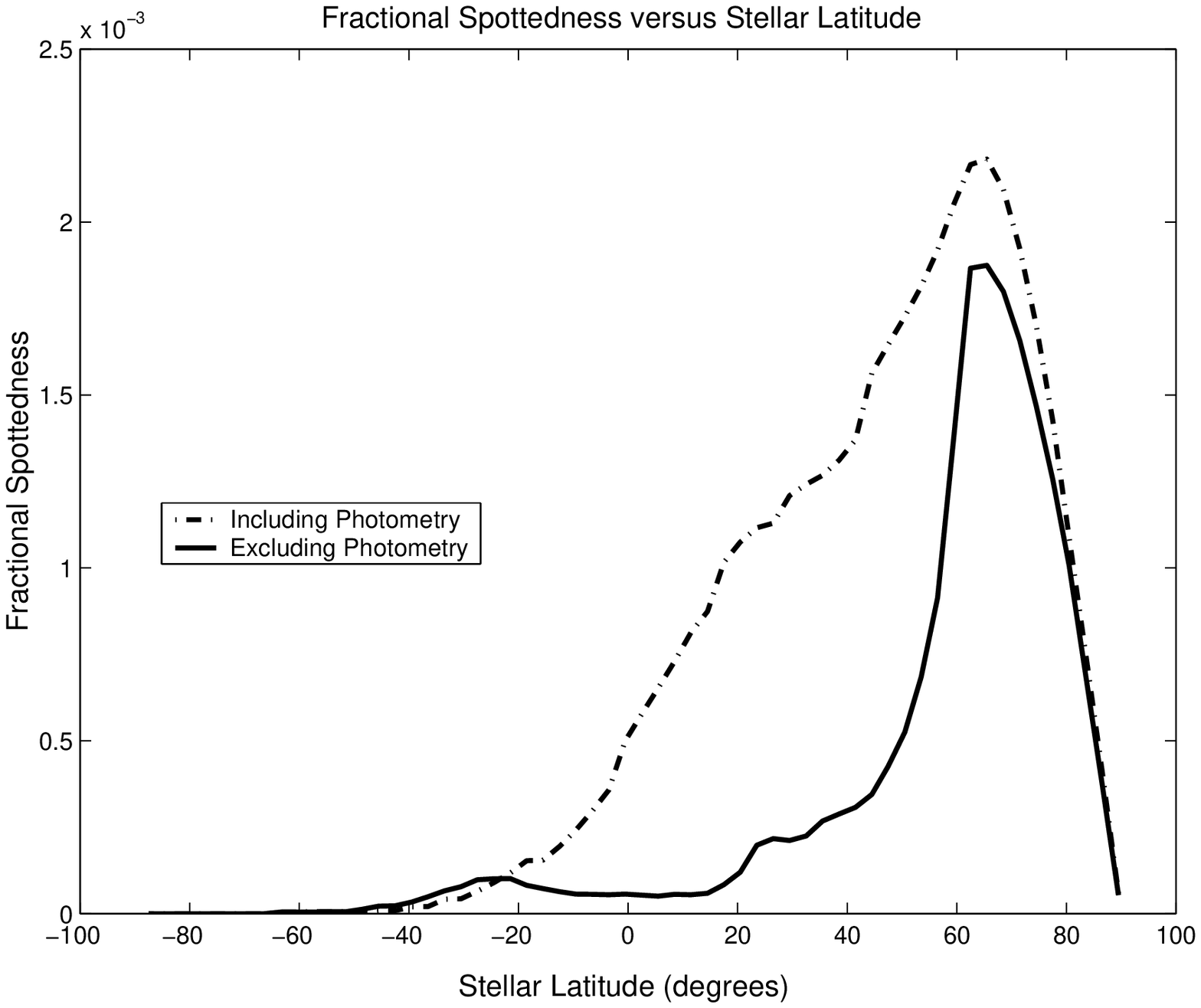}
\caption{The fractional spottedness versus stellar latitude for HD106506 with and without photometry included in the imaging process. The fractional spottedness is based on the average spot occupancy at each longitude and is defined in equation~\ref{spottedness}. The photometry is very useful at extracting out the lower latitude features whereas the spectroscopy extracts the higher latitude features.}.
\label{fractional_spottedness}
\end{center}
\end{figure}

Spectroscopic and photometric observations were used to produce the spot occupancy map shown in Figure \ref{Photometry_incl}. Photometry is very useful at extracting out the lower latitude features (Jeffers, Barnes \& Collier Cameron, 2002) but is not sensitive to the high-latitude features as these are always in view, whereas the spectroscopy extracts the higher latitude features \citep{Unruh97} but cannot discriminate lower latitude features at latitudes $\la$ 30 degrees. Hence to get a more complete picture of the stellar image, both photometric and spectroscopic data need to be incorporated into the imaging process. Using the Stokes {\it I}  spectroscopic information coupled with the photometry, the spot occupancy map, as shown in Figure \ref{Photometry_incl} show low-latitude to mid-latitude spots, and the predominant polar spot. This variation of spot occupancy with stellar latitude using spectroscopy alone and inclusive of the photometry is shown in Figure \ref{fractional_spottedness}. The fractional spottedness is defined in equation~\ref{spottedness}.

\begin{equation}
	\label{spottedness}	
	F(\theta) = S(\theta)cos(\theta)d\theta /2
\end{equation}
where F($\theta$) is the fractional spottedness at latitude $\theta$, S($\theta$) is the average spot occupancy at latitude $\theta$, and d$\theta$ is the latitude width of each latitude ring.

The exact reasons why rapidly rotating stars have polar spots are still not fully understood. One theory \citep{Schussler96, Buzasi97, DeLuca97} postulates that high-latitude, but not truly polar, spots are a result of the increased Coriolis effect, due to the rapid rotation, which may have the effect of deflecting the spot features to higher latitudes as they erupt through the star's convection zone. An alternative hypothesis \citep{Schrijver00, Kitchatinov99} involves the transport of flux through the process of meridional flow. It is yet to be determined whether the polar spots on young solar-type stars are formed at high latitudes by the strong Coriolis effect in these rapidly rotating stars, or formed at low latitudes and pushed poleward by subsurface meridional flows. Nevertheless, the issue can be clarified by long-term monitoring of the spot topology of these stars, because meridional flows should be detectable as a poleward drifts in starspot features over time. \citet{Weber05} have reported tentative evidence for large poleward meridional flows on early-K giants, but to the best of our knowledge, such an effect has yet to be detected for young solar-type stars.

These reconstructed images are consistent with those obtained in other Doppler images produced for young, single, rapidly rotating solar-type stars. The photometry enabled more of the low- latitude features to be recovered, as was done for the star R58 (in IC 2602) by \citet{Marsden05}.

\begin{figure}
\begin{center}
\includegraphics[scale=0.45, angle=0.0]{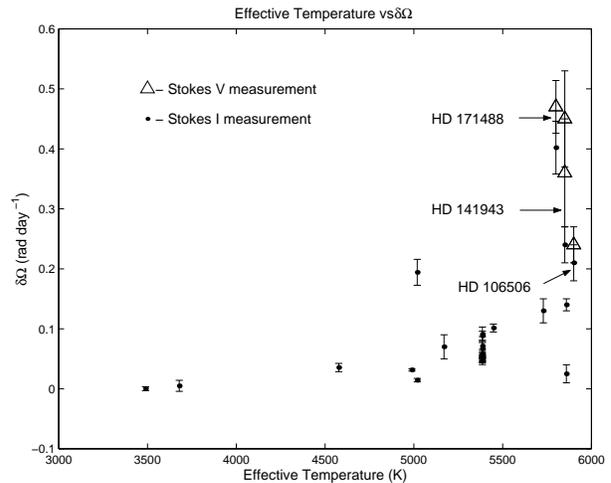}
\caption{This shows a graph of $\delta\Omega$ versus effective surface temperature, as orginally produced by \citet{Barnes05}. The data for HD~171488 \citep{Marsden06}, HD~141943 \citep{Marsden10a} and HD~106506 have been added to this graph. HD~106506 supports the power law developed by \citet{Barnes05} whereas HD~171488 and HD~141943 appears not to support this relationship.}
\label{convectiveZone}
\end{center}
\end{figure}

When using the Stokes $\it{I}$ information HD~106506 was found to have a photospheric shear of $0.21_{-0.03}^{+0.02}$ \rdd with the equator lapping the poles every $\sim$ $30_{-3}^{+5}$ days. When compared with the Sun's value for $\delta\Omega$=0.055 \rdd, with a lap time of 115 days, this young pre-main sequence solar-type star has a shear of roughly 4 times greater than the Sun. This result is consistent with the relationship determined by \citet{Barnes05} that differential rotation is dependent on spectral class (surface temperature), with M-dwarfs rotating almost as a solid body through to early-G dwarfs exhibiting a strong rotational shear. However, \citet{Marsden06} and \citet{Jeffers08} have found that the early G-dwarf, HD 171488 (V889 Her) has a photospheric shear up to seven times the solar value. So why do these two stars that exhibit similar surface temperature properties have vastly different photospheric shear? One possible reason is that HD~106506 is a more distended star, with a larger convective zone when compared with HD~171488. Being a younger star, it is probably still contracting down to the main sequence. But it is still unclear why HD~171488 should have such a large d$\Omega$ and perhaps it is convective zone depth and the rate at which that zone is turned over that is more important than surface temperature (hence spectal type). Figure \ref{convectiveZone} is a graph of $\delta\Omega$ versus effective temperature, as orginally produced by \citet{Barnes05} with HD~106506, HD~141943 \citep{Marsden10b} and HD~171488 \citep{Marsden06,Jeffers08} added to the dataset. HD~106506 supports the power law developed by \citet{Barnes05}.

\begin{figure}
\begin{center}
\includegraphics[scale=0.45, angle=0.0]{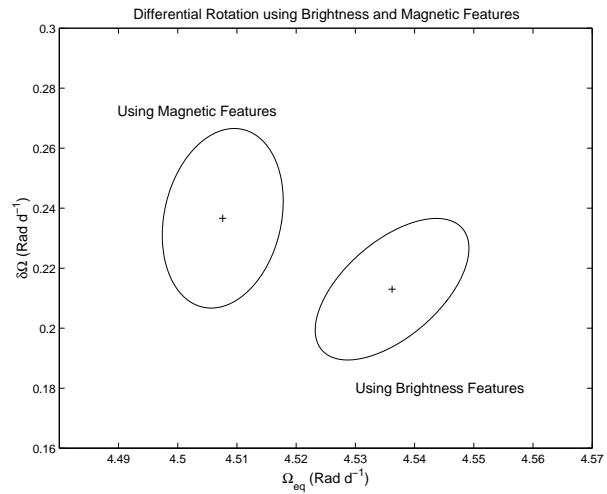}
\caption{This shows a graph of the error ellipses for the differential rotation measurements for both the Stokes $\it{I}$ (on the right) and the Stokes $\it{V}$ (left). These error ellipses were generated by varying some of the stellar parameters, including the star's inclination ($\pm$ 10$^{o}$) and the Stokes $\it{I}$ spot coverage or the Stokes $\it{V}$ global magnetic field ($\pm$ 10\%). }
\label{error_ellipses}
\end{center}
\end{figure}

\begin{table*}
\begin{center}
\caption{A comparison with other late F/early G type dwarfs that have had their surface differential rotation measured using Doppler imaging. } 
\label{HD171488_comp}
\begin{tabular}{llllll}
\hline
Parameter  		& HD 106506	     & HD 171488 $^{2,3}$     & R58 (in IC2602) $^4$ & LQ Lup $^5$     & HD 141943$^6$	\\
\hline
(B-V)$^1$		& 0.605		     & 0.62		      & 0.61		 & 0.69 	   & 0.7  \\
T\subs{Photosphere} K	& 5900  	     & 5800                   & 5800 $\pm$ 100     & 5750 $\pm$ 50     & 5850 $\pm$ 100 	\\
T\subs{Spot} K		& 4000 	             & 4200 		      & 3900 $\pm$ 100     & 4250	           & 3950 \\
Age (Myr) 		& $\la$10  	     & 30-50                  & 35 $\pm$ 5         & 25 $\pm$ 10       & 17 \\ 
Mass $M_{\odot}$	& 1.5 $\pm$ 0.1       & 1.06 $\pm$ 0.02          & 1.15 $\pm$ 0.02    & 1.16 $\pm$ 0.02   &  1.3 \\
Radius $R_{\odot}$	& 2.15 $\pm$ 0.26      & 1.15 $\pm$ 0.08          & 1.18 $\pm$ 0.02    & 1.22 $\pm$ 0.02   &  1.6 $\pm$ 0.1\\
Incl. Angle 		& 65 $\pm$ 5$^{o}$     & 60 $\pm$ 5$^{o}$         & 60 $\pm$ 10$^{o}$  & 35 $\pm$ 5$^{o}$  &  70 $\pm$ 10   \\
\vsini  \kmsn		& 79.5 $\pm$ 0.5       & 37.5 $\pm$ 0.1           & 92 $\pm$ 0.5  	 & 115 $\pm$ 1       &  35.0 $\pm$ 0.5		\\
Stokes $\it{I}$ Measurements   & & & & & \\
$\Omega$\subs{eq}  \rdd & 4.54 $\pm$ 0.01      & 4.786 $\pm$ 0.013        & 11.165 $\pm$ 0.014 &  20.28 $\pm$0.01 & 2.86 $\pm$ 0.02  \\
$\delta\Omega$  \rdd	        & $0.21_{-0.03}^{+0.02}$      & 0.402 $\pm$ 0.044        & 0.0815 $\pm$ 0.026 & 0.12 $\pm$0.022  & 0.24 $\pm$ 0.03 \\
Stokes $\it{V}$ Measurements   & & & & & \\
$\Omega$\subs{eq}  \rdd & 4.51 $\pm$ 0.01      & 4.85 $\pm$ 0.05          & -- & --& 2.88 $\pm$ 0.02 $^7$ \\ 
			& 		     & 			      &    &   & 2.89 $\pm$ 0.05 $^7$ \\
$\delta\Omega$  \rdd	        & 0.24 $\pm$ 0.03      & 0.47 $\pm$ 0.04 	      & -- & --& 0.36 $\pm$ 0.09 $^7$ \\
			& 		     & 			      &    &   & 0.45 $\pm$ 0.08 $^7$ \\
\hline
\end{tabular}
\end{center}
$^1$ Hipparcos, $^2$ \citet{Marsden06}, $^3$ \citet{Jeffers08}, \\
$^4$ \citet{Marsden05}, $^5$ \citet{Donati00}, $^6$ \citet{Marsden10a} \\
$^7$ \citet{Marsden10b}\\
\end{table*}

\subsection{Magnetic Topology}

The reconstructed magnetic fields are shown in Figure \ref{mapmag} and the fits to the Stokes {\it V} LSD profiles are given in Figure \ref{mag_fits}. Magnetic features in both the azimuthal and radial magnetic fields extend from low-latitude to high-latitude features. As discussed by \citet{Donati03a}, the radial component may be interpreted to be representative of the poloidal component while the azimuthal component may be considered as representive of the toroidal component of the large-scale dynamo field. The strong azimuthal ring around the star is similar to those observed on other stars such as HD 171488 \citep{Jeffers08}. However, the azimuthal ring on HD~106506 is of different polarity to those observed on other stars, such as HD~141943 \citep{Marsden10a}. This is unlikely to be due to instrumental polarisation as the data for HD~106506 were taken with the same instrumental setup and at the same time as HD~141943.

The azimuthal ring of magnetic field is consistent with the wreathes of magnetic fields theorised by \citet{Brown10}, except at a much higher latitude. Their 3-dimensional MHD models found that stars with increased rotation rate (up to 3 times that of the Sun) could produce self-generating wreathes of azimuthal field without the need for an interface layer. One can speculate that the increased rotation rate of HD 106506 could be deflecting the wreathes of magnetic fields to much higher latitudes, as seen in Figure 7.

\citet{Donati09} found that increasing rotation rate of a star, the toroidal component of the surface magnetic field dominates over the poloidal component of that field. The magnetic field on HD~106506 is predominantly toroidal with approximately 70 $\pm$ 3\% of the magnetic energy being toroidal magnetic energy. This is similar with the toroidal field being the dominant field of HD~171488 \citep{Marsden06,Jeffers08}, AB Dor and LQ Hya \citep{Donati03a}. Like HD~171488, this field is quite complex and much more than that operating the Sun. Considering the toroidal component, an octopole field dominates with approximately 37 $\pm$ 9\% of the magnetic energy whereas the dipole field contains only 18 $\pm$ 5\% and a quadrople field containing approximately 16 $\pm$ 7\% of the magnetic energy. This is in contrast to the poloidal field were the respective components are roughly equal between a dipole (10 $\pm$ 2\%), quadopole (11 $\pm$ 2\%) and octopole field (10 $\pm$ 1\%). This would indicate that a much more complex dynamo is operating on HD~106506 and is similar to that on other young, sun-like stars. In addition, the toroidal field was predominantly axisymetric (95 $\pm$ 2\%) whereas the poloidal field tended to be more non-axisymetric (65 $\pm$ 7\%). This is consistent with the observations of \citet{Donati09} and \citet{Marsden10a} where stars with significant toroidal fields often have non-axisymetic poloidal fields. The error estimates on these measures were obtained by varying a range of stellar parameters such as inclination, radial velocity, $\Omega$\subs{eq}, $\delta\Omega$, period and \vsini. Hence these may be refered to as a $\it variation$ bar, which is analogous to an error bar except that the measurement indicates the variation found by this method.

When measuring the differential rotation using the magnetic features, the equatorial rotation rate $\Omega$\subs{eq} was 4.51 $\pm$ 0.01 \rdd, with a photospheric shear $\delta\Omega$ of 0.24 $\pm$ 0.03 \rdd. The comparison with those parameters found using the Stokes $\it{I}$ information is shown in Figure \ref{error_ellipses}. However, this difference in the $\delta\Omega$ values are within the error bars of the two measurements. When compared with the differences observed on other stars of similar spectral type such as HD~141943 \citep{Marsden10b}, these features may be anchored at similar depths in the convective zone, unlike HD~141943 which may be anchored at different depths of the convective zone. However, this conclusion must be treated with caution until further Stokes $\it{V}$ data is obtained for this star.

\subsection{H$\alpha$ emission}

Figure \ref{chromo1} shows a bright spot in this dynamic profile at $\sim$0.15 phase. This is where the chromosphere is much denser as a result of an increase in temperature in that local region. There appears to be a reduced emission at $\sim$0.75 phase as well. These variations in H$\alpha$ emission lie within the stellar velocity profile and hence arise from regions close to the stellar surface. We thus interpret these H$\alpha$ variations in terms of prominences more akin to solar prominences than the more extended features knows to exist at the co-rotation radius of other active solar-type stars, such as LQ Lup \citep{Donati00}.

\section[Conclusions]{Conclusions}

In this paper, reconstructed brightness and magnetic images of the young solar-type star HD 106506 have been presented. The brightness image shows low-latitude to mid-latitude features coupled with a predominate polar spot. This is very similiar to many of rapidly rotating solar-type stars. The magnetic images reveal regions of azimuthal field near the stellar surface that suggest that the dynamo mechanism may be occuring entirely in the convective zone and close to the surface of the star itself.  The photospheric shear for HD 106506 supports the findings of \citet{Barnes05}, that effective temperature is the dominant factor in the level of surface differential rotation. However, the reason why such a relation exists remains unclear.

\section*{Acknowledgments}

The authors would like to thank the technical staff at the AAT for their helpful assistance during the acquisition of the spectroscopic data. Also to Dr John Kielkopf (UofL) for his expertise in developing the University of Louisville Telescope at MKO and to Roger McQueen in taking some of the photometric data. We thank the anonymous referee for their diligence and many helpful comments that have made substantial improvements to this paper. This project is supported by the Commonwealth of Australia under the International Science Linkages program. This project used the facilities of SIMBAD, HIPPARCOS and IRAF. This research has made use of NASA's Astrophysics Data System.

\appendix
\section{BVR Photometry at MKO: 2005 and 2007}

The IRAF package was used for the reduction and analysis of the data. The CCD frames had the dark current removed (that also contained the Bias information) and were flat fielded using a nightly master flat field, which was a combination of 9 dome flat fields. To produce a single data point, three images taken through each filter of each star field were averaged, using IMCOMBINE, to produce an image with improved signal-to-noise from which the magnitude of all the program stars were measured. The stars in the frame were found using DAOFIND and simple aperture photometry was done using DAOPHOT. The sky background was determined and removed using the mode of an annulus of 10 pixel radius around the star. The curves of growth method \citep{Stetson90} was used to determine the aperture size that would be used to determine the magnitude of each of the stars in the field. 

\subsection{Standardisation of Comparison Stars: 25th July, 2005}

BVR photometry was conducted at MKO in 2005 in order to determine the photometric colour of HD 106506 and the other nearby stars within the field of view. Figure \ref{phot_image} shows the target star and 5 comparison stars within the same field of view, are indicated. Several BVR photometric observations were taken of the standard stars in Graham's E-regions \citep{Graham82}. The O'Mara telescope used for this procedure comprises a 35cm Schmidt-Cassegrain optics, an SBIG STL-1301E CCD camera with 1280x1014 16 $\mu$m square pixels, and a Paramount ME mount. The three regions were used, the E-5 (Stars A,O,U,S,V,Y,c), E-6 (Stars 98,M.P,X,W) and E-7 (Stars M and S) regions. The routines found in the PHOTCAL package in IRAF were used to determine both the transformation coefficients and the magnitude of the stars, including HD~106506. There were three images taken through each filter of the star fields and then averaged using IMCOMBINE, to produce a single image from which the magnitude of all the standard stars and the program stars were measured. The following transformation equations were solved using the FITPARAM task in IRAF:

\begin{figure}
\begin{center}
\includegraphics[scale=0.47, angle=0]{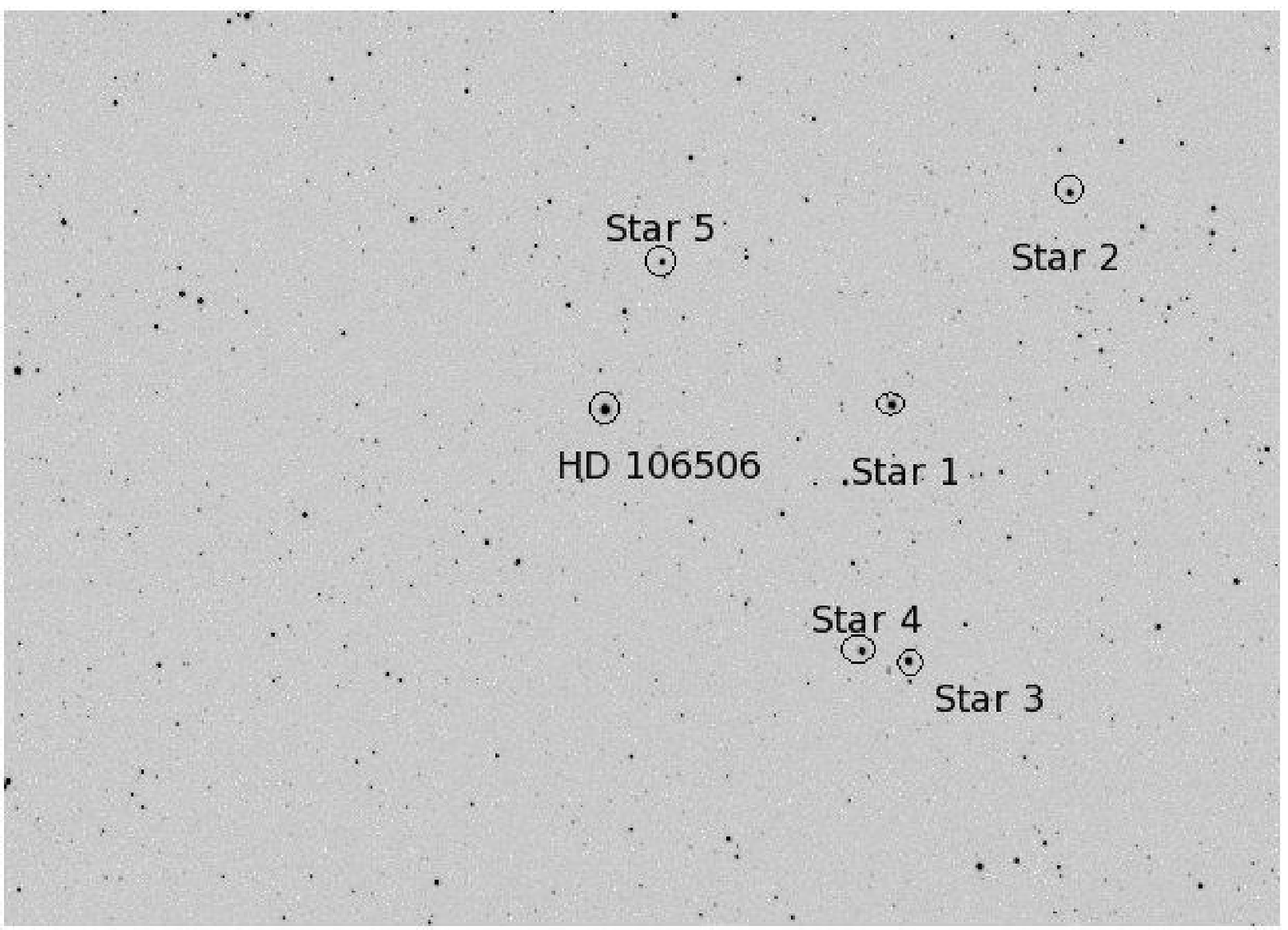}
\caption{The area surrounding HD 106506, showing 5 of the brighter stars were used in the determination of the period. The stars' names are shown in Table \ref{bvri_phot} in appendix A.} 
\label{phot_image} 
\end{center}
\end{figure}

\begin{table}
\begin{center}
\caption{Transformation Coefficients for the 25th July, 2005 photometry at MKO. The associated errors, rms fit and $\chi$\sups{2} value for each filter have been included} 
\label{coefficients}
\begin{tabular}{lccccc}
\hline
n & $b_{\rm n}$ & $v_{\rm n}$ & $r_{\rm n}$ \\
\hline
0 &  -3.217     & -3.506      & -3.709      \\
  & $\pm$0.14 &$\pm$0.13  &$\pm$0.12  \\

1 &  0.311      & 0.144       &  0.134      \\
  & $\pm$0.11 &$\pm$0.09  &$\pm$0.10  \\
2 &    -0.154   & 0.065       & 0.020       \\
  &$\pm$0.12  &$\pm$0.06  &$\pm$0.14  \\
\hline
$\chi$\sups{2}  & 0.103   &  0.060 & 0.023  \\
rms fit    & 0.021   & 0.019  & 0.011       \\
\hline
\end{tabular}
\end{center}
\end{table}

\begin{table}
\begin{center}
\caption{The BVRI information for HD 106506 and the comparison stars. \citet{Graham82} E-Regions, E5,E6 and E7, were used to standardise these stars and the data was processed using the routines in PHOTCAL found in IRAF. The data were taken on the 25th July, 2005 and a sample image, with the star number included, is shown in Figure \ref{phot_image}.}
\label{bvri_phot}
\begin{tabular}{llccc}
\hline
Star & Star's Name & V & BV & VR  \\
\hline
Target    & HD 106506       &  8.463   &  0.594  & 0.356   \\
Star 1    & HD 106507       &  9.152   & -0.024  & 0.044   \\
Star 2    & TYC 8979-979-1  & 10.108   &  0.100  & 0.101   \\
Star 3    & TYC 8978-1180-1 & 10.003   &  1.083  & 0.584   \\
Star 4    & HD 106417       & 10.133   &  1.015  & 0.558   \\
Star 5    & TYC 8978-5479-1 & 10.948   &  0.203  & 0.165   \\
\hline
\end{tabular}
\end{center}
\end{table}

\begin{equation}
	\label{b_phot}
	{m_{\rm b} = B + b_{\rm 0} + b_{\rm 1}X_{\rm b} + b_{\rm 2}(B-V)} 
\end{equation}

\begin{equation}
	\label{v_phot}
	{m_{\rm v} = V + v_{\rm 0} + v_{\rm 1}X_{\rm v} + v_{\rm 2}(B-V)} 
\end{equation}

\begin{equation}
	\label{r_phot}
	{m_{\rm r} = R + r_{\rm 0} + r_{\rm 1}X_{\rm r} + r_{\rm 2}(V-R)} 
\end{equation}

where B, V and R are listed magnitudes of the standard stars in \citet{Graham82}. The instrumental magnitudes, as measured by DAOPHOT were $m_{\rm b}, m_{\rm v}$ and $m_{\rm r}$. The zeropoints are $b_{\rm 0},v_{\rm 0}$ and $r_{\rm 0}$; while $b_{\rm 1},v_{\rm 1}$ and $r_{\rm 1}$ are the extinction coefficients and $b_{\rm 2},v_{\rm 2}$ and $r_{\rm 2}$  are the colour correction coefficients. The values determined for the coefficients are given in Table \ref{coefficients} and the results are shown in Table \ref{bvri_phot}.

\subsection{BVR Broadband Photometric Observations from MKO: 2007}

Table \ref{photometry_log} shows a log of data taken during the observation run in 2007. The aperture size selected had a pixel radius of 15, giving an aperture radius of approximately 8 arc seconds. On the 11th of April, several observations were taken of the standard stars in Graham's E5-regions \citep{Graham82} so as to place HD 106506 on the standard magnitude scale. The stars used were A,O,U,S. The routines found in the PHOTCAL package in IRAF were used to determine both the transformation coefficients and the magnitude of the stars, including HD 106506. There were three images taken through each filter of each star field and these were averaged, using IMCOMBINE, to produce an image with improved signal-to-noise from which the magnitude of all the standard stars and the program stars were measured. 

\subsection{Period Determination in 2007}

\begin{figure}
\begin{center}
\includegraphics[scale=0.48, angle=0]{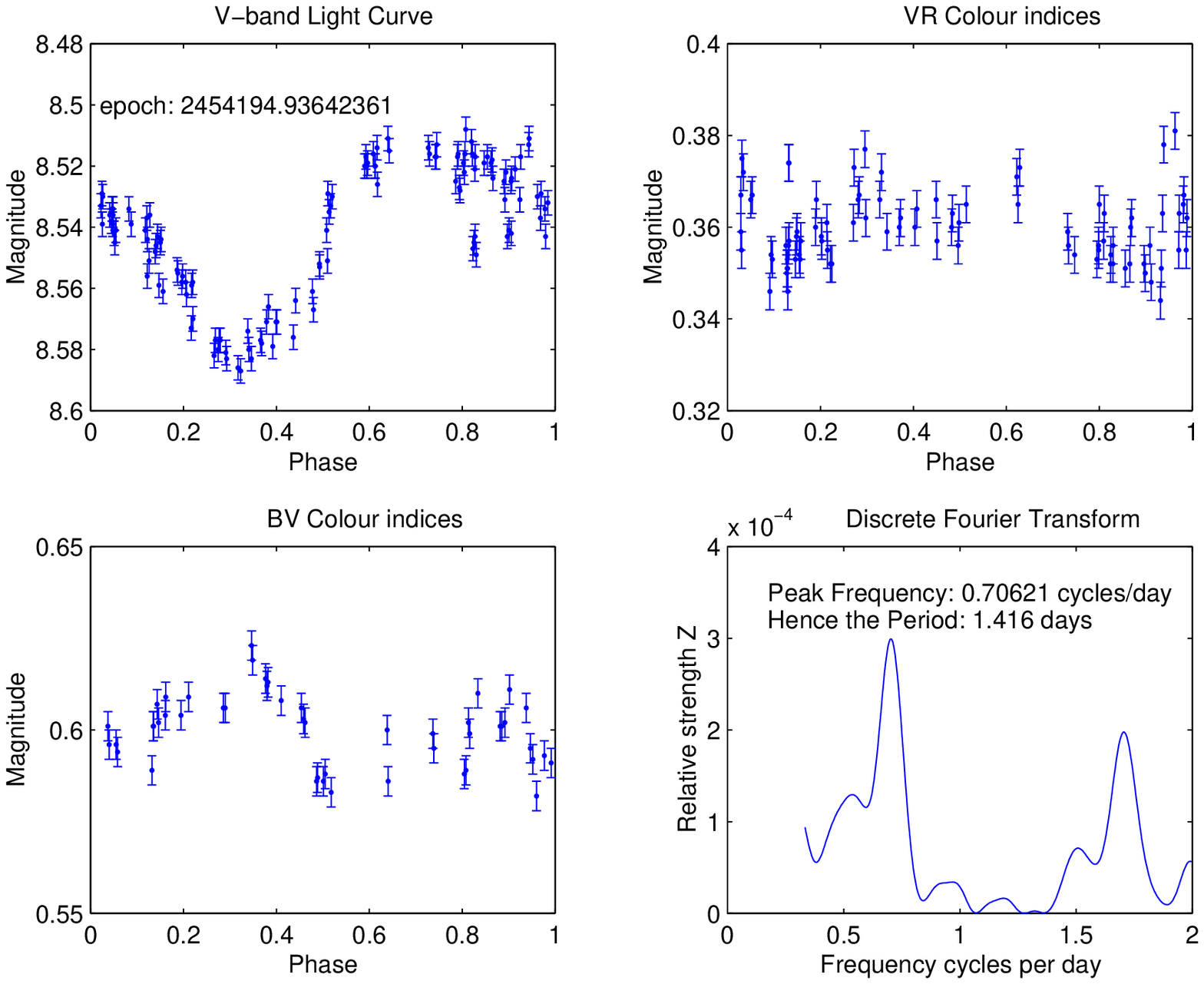}
\caption{The top left panel shows the V-band light curve for HD 106506, with the phase referenced to epoch = 2453543.93642361-d which coincided approximately with the middle of the AAT run. The BV and VR colour light curves are also shown. The bottom right panel shows the Discrete Fourier transform, which is interpreted in terms of a rotation period of 1.416 days.}
\label{period_dft} 
\end{center}
\end{figure}

The rotational period was determined using Discrete Fourier transforms (DFT), using the algorithm as explained by \cite{Berlesene88}. The five brighter stars were selected as possible comparison stars and the period was determined for each in B, V and R band. The average period was determined to be 1.416 $\pm$ 0.133 days, and was consistent with all the comparison stars in B, V and R bands. The error was determined by finding the full width half maximum of the DFT line profile. Figure \ref{period_dft} shows the phased folded light curve for the V- band as well as the BV and VR colour indices. The power spectrum is also shown as generated by the DFT algorithm. Graham's E5 Region, coupled with the 5 stars previously standardised,  were used to place HD~106506 onto the Cousins magnitude system. 

\bsp

\label{lastpage}

\end{document}